# INFORMATION-CONTROL SOFTWARE FOR HANDLING SERIAL DEVICES IN AN EPICS ENVIRONMENT*

P. Chevtsov, S. Schaffner, Jefferson Lab, Newport News, VA 23606, USA

*Abstract*

Each accelerator control system has a variety of measurement devices. One of the most common types of instrument interfaces used for their control is a serial (RS-232) bus. It is inexpensive and adequate for relatively simple measurement and control devices such as switchers, amplifiers, voltmeters, and stepper motors. Since the RS-232 specification is very broad and does not require uniformity above the basic communication protocol level, one of the major problems associated with the use of RS-232 is that the command protocol for each device is unique. This makes it difficult to design generic drivers for RS-232 and also hampers efforts to develop generic troubleshooting methods. This paper presents software developed independently at three other labs and integrated into a single system at Jefferson Lab to handle serial devices in a generic manner. The software is based on the EPICS toolkit [1] and uses a 3-tier architecture including a common serial driver at the bottom, a top-level protocol to specify individual device commands in a generic manner, and mid-level software to "glue" the two together.

## 1 INTRODUCTION

RS-232 is one of the oldest computer communication standards. It was introduced in the early 1960s as the recommended standard (RS) number 232 for the "Interface between Data Terminal Equipment and Data Communications Equipment Employing Serial Binary Data Interchange" [2]. There are two types of serial devices: the Data Terminal Equipment (DTE) and Data Communication Equipment (DCE). The computer is considered a DTE, while peripheral devices, such as modems, are DCEs. The two are linked together through a serial port and the connection is relatively slow since data travels across a wire one bit at a time. Serial ports consist of two signal types: data signals and control signals. The physical connection in a serial line can be as simple as three wires for "data in", "data out" and signal ground. Additional wires may be used for "handshaking" when two ends of the communication line tell each other that they are ready to receive data. RS-232 also defines a communication method to send information on a physical channel. The information is broken up in data words. The length of the data word is variable, usually 7 or 8 bits. Data bits are sent with a predefined frequency, the baud rate that is roughly equivalent to the number of bits transmitted per second. To understand each other, a transmitter and a receiver must use the same number of bits in a data word and the same baud rate.

The RS-232 standard says nothing about the content of the message sent between the two devices, the amount of time to wait between messages and how to handle multi-word messages. These issues are left up to the designer of the device. The result is that each device is unique in the way messages are interpreted at both the DTE and DCE ends and this makes it difficult to develop a generic method of programming that will support the range of possible RS-232 devices.

## 2 DEVELOPMENT, DIAGNOSTIC AND INFORMATION TOOL FOR RS-232

### 2.1 Background

The accelerator control system at Jefferson Lab is based on the EPICS toolkit. With its very powerful set of tools, EPICS allows easy system extensions at all control levels. Several devices within the control system utilize serial communications. Most of these systems connect to the serial ports on the Input-Output Controller (IOC) transition module. Since it is relatively easy to access these serial ports using the standard terminal I/O driver supplied with VxWorks, such systems treat the serial ports as file I/O and write custom EPICS subroutine records or State Notation Language (SNL) sequencers to control devices through simple read/write commands. This is not an optimal way to handle device control for several reasons. First, there are only a very limited number of available ports on the transition modules (usually, 2-4). Second, this method bypasses the standard EPICS model that divides device control into three distinct layers: a layer which interfaces directly with the device (driver support) and is essentially independent of EPICS, a layer which handles communications with Channel

---

* Supported by DOE Contract #DE-AC05-84ER40150

Access, the EPICS network protocol, and has no real knowledge of the underlying device (record support), and a layer in between that handles the interface between the two (device support). The method that has been in use at Jefferson Lab moves all of the processing into the top layer. This makes it difficult to properly identify serial applications since hardware identification in EPICS is handled through the device layer. It also makes such tasks as upgrading the version of EPICS or changing processor type difficult since each of these serial applications shares no common device or driver support and each must be upgraded, tested and verified separately.

The serial device control at Jefferson Lab followed this non-standard pathway because EPICS is based on an inherently register-based control algorithm. Each record assumes that it is connected to a single control channel providing the same type of information with every access. Serial devices can serve a variety of data types and command sequences through a single channel. This makes it difficult to fit into the EPICS model. A detailed analysis of many serial control applications that have been created in the EPICS community reveals that all the elements for a generic EPICS-based serial control model exist. All that is needed is to put them together to form a single system.

RS-232 was one of the first subnets supported by EPICS. An EPICS terminal driver and low level drivers for several common serial interface devices were developed [3]. Many attempts were made to fashion traditional EPICS device support on top of this layer but these attempts were not successful for the reasons mentioned above. Two recent additions to the EPICS code base have provided the missing pieces. The first is the Message Passing Facility (MPF) [4] developed by Marty Kraimer at Argonne National Laboratory. The second is the Stream Device Support software [5] developed by Dirk Zimoch at the University of Dortmund. The MPF system was developed as a device support layer mechanism to pass messages to and from processes. It does not directly connect to EPICS records. The Stream Device Support software connects directly to EPICS records but does not have an interface to the serial driver software. The tool developed at Jefferson Lab combines these two device support layers along with several serial drivers to provide configurable, generic serial support software. The tool consists of the serial device/driver support library, an information system about the use of this library for various control applications, and the details of the RS-232 standard.

The serial device/driver support library uses a three tier software architecture: a common serial driver at the bottom, a top-level protocol to specify individual device commands in a generic manner, and a mid-level software to "glue" the two together.

## 2.2 Common Serial Driver

This tier deals with the serial ports used for the communication with control devices. It is multi-component software. Each type of supported RS-232 communication hardware is handled by the corresponding software component. Currently, the VME serial transition modules, "on board" IOC serial ports, IP-Octal modules, and 16-channel intelligent asynchronous serial controllers VMIVME-6016 are supported. It should be fairly easy to integrate other serial drivers into the system.

Each serial port is served by a separate control task that exchanges messages with the EPICS database records referencing this port. The serial port control task reads and writes data into and out of the serial port. It also handles the hardware operation timeouts.

The driver layer is activated with the use of two basic calls. The first call initializes the serial communication hardware. The second call sets the basic communication parameters for each serial port provided by the underlying hardware: the baud rate, data word size, number of stop bits, parity, etc. All activation calls for each IOC are combined into one serial configuration file. This file is downloaded into the IOC at start up time and activates the driver. The control system is then ready to communicate with serial devices connected to the initialized serial ports.

## 2.3 Serial Protocol Software

Each type of serial control device has its own set of commands and data formats. Every action of the device consists of a sequence of input and output operations and a set of additional parameters that affect data transfer, such as separators, terminators, or timeouts. This is called the serial protocol. Each serial protocol has a name that is usually associated with a particular device action. All serial protocols for each device type are combined in one serial protocol file. All serial protocol files reside in one directory. The file name, the protocol name, and the information about the type of RS-232 hardware and serial port on this hardware are referenced by INP or OUT links of EPICS database records. At IOC initialisation time, the serial protocol software parses the protocol file and defines the device control actions for all such records. This information becomes part of these EPICS records.

## 2.4 Middle-Layer Software

This software layer is used to "glue" the two tiers described above and to provide a user with RS-232 diagnostics.

The data communication part of this software is the interface between the EPICS database and the serial driver. The data communication software accepts the requests from EPICS records, packs them into the messages with the data conversion specified by the protocol, sends these messages to the corresponding serial port control task, waits for the responses from this control task, makes the necessary data conversion, and informs the records of the results of the performed control operations.

The diagnostic part of this software is a set of service utilities continuously monitoring the serial ports and generating reports about all data streams resulting from user requests. The information system of the serial development, diagnostic and information software tool uses the link to our RS-232 Management Web Page. It has the description of the serial device/driver support library and many examples of its use at Jefferson Lab to control various serial devices. It also contains the RS-232 standard details as well as very effective troubleshooting procedures. The diagnostic software and information system are very useful in the event that serial communication problems arise.

# 3 IMPLEMENTATION OF THE SERIAL DEVELOPMENT TOOL

One of the main advantages of our serial development tool is that it typically does not require any software coding for connecting a new serial device to the control system. All that needs to be done to set up EPICS controls for a serial device using one of the supported drivers is:

**a)** read the device manual to find out the details of the command protocol and describe it in a serial protocol file (following numerous examples from the information system);
**b)** determine serial hardware that will be used (IPAC, transition module, etc.), find out from the device manual the communication parameters of the serial port to which the device will be connected, and specify all this in the IOC serial hardware configuration file;
**c)** create the EPICS database that will handle the serial device command protocol;
**d)** connect the device to the serial communication port with a proper cable;
**e)** initialize serial control and diagnostic software for the serial port with the use of the serial hardware configuration file mentioned above;
**f)** run the EPICS database to control the device;
**g)** use the diagnostic software and information system for troubleshooting in case of serial communication problems.

The serial device/driver support library has been successfully tested in a few systems at Jefferson Lab. The first of these systems was finished in April 2001. Fifteen serial devices (steppermotors, temperature controllers, lasers) used in the He3 Polarized Target experiment [6] were completely controlled by one IOC (MVME-162 board). The serial device/driver support library ran on this computer together with many other applications such as laser and steppermotor interlock programs. The system has worked smoothly for 24 hours a day, 7 days a week for several months now.

# 4 CONCLUSION

Our serial development, diagnostic and information software tool significantly simplifies the process of the creation of control applications for serial devices. In most cases, it does not require software coding. It contains the information system and diagnostic software providing generic troubleshooting methods for RS-232. We plan to slowly migrate our current serial devices to this system and integrate GPIB, another message passing protocol, into this system.

# REFERENCES

[1] L.R. Dalesio et al., "The Experimental Physics and Industrial Control System Architecture: past, present and future", NIM, A 352 (1994).
[2] J. Axelson, "Serial Port Complete", 359 p.
[3] A.Johnson, P. McGehee, A. Honey, J. Hill et al. http://www.slac.stanford.edu/grp/cd/soft/epics/site/tyGSOctal.html
[4] M. Kraimer, "Message Passing Facility" http://www.aps.anl.gov/asd/people/mrk/epics/modules/bus/mpf
[5] D. Schirmer et al. "Standardization of the DELTA Control System", ICALEPCS-99, p. 75-77.
[6] J.-P.Chen "He3 Polarized Target Collaboration" http://hallaweb.jlab.org/physics/experiments/he3/